# Ultra-sensitive graphene-based electro-optic sensors for optically-multiplexed neural recording

Zabir Ahmed[1], Xiang Li[1], Kanika Sarna[1], Harshvardhan Gupta[1], Vishal Jain[1,2], Maysamreza Chamanzar[1,2,3]

1. Department of Electrical and Computer Engineering, Carnegie Mellon University, Pittsburgh, PA, 15213, USA.
2. Carnegie Mellon Neuroscience Institute, Pittsburgh, PA, 15213, USA.
3. Department of Biomedical Engineering, Carnegie Mellon University, Pittsburgh, PA, 15213, USA.

## Abstract

Large-scale neural recording with high spatio-temporal resolution is essential for understanding information processing in brain, yet current neural interfaces fall far short of comprehensively capturing brain activity due to extremely high neuronal density and limited scalability. Although recent advances have miniaturized neural probes and increased channel density, fundamental design constraints still prevent dramatic scaling of simultaneously recorded channels. To address this limitation, we introduce a novel electro-optic sensor that directly converts ultra-low-amplitude neural electrical signals into optical signals with high signal-to-noise ratio. By leveraging the ultra-high bandwidth and intrinsic multiplexing capability of light, this approach offers a scalable path toward massively parallel neural recording beyond the limits of traditional electrical interfaces. The sensor integrates an on-chip photonic microresonator with a graphene layer, enabling direct detection of neural signals without genetically encoded optical indicators or tissue modification, making it suitable for human translation. Neural signals are locally transduced into amplified optical modulations and transmitted through on-chip waveguides, enabling interference-free recording without bulky electromagnetic shielding. Arrays of wavelength-selective sensors can be multiplexed on a single bus waveguide using wavelength-division multiplexing (WDM), greatly improving scalability while maintaining a minimal footprint to reduce tissue damage. We demonstrate detection of evoked neural signals as small as 25 µV with 3 dB SNR from mouse brain tissue and show multiplexed recording from 10 sensors on a single waveguide. These results establish a proof-of-concept for optically multiplexed neural recording and point toward scalable, high-density neural interfaces for neurological research and clinical applications.

## Introduction

Unraveling the intricate dynamics of neural circuits that mediate the brain function requires tools capable of accurately mapping neuronal activity across various brain regions with high spatial and temporal precision, all while being minimally invasive. Neural circuits are densely packed with neurons; for instance, the mouse cortex contains nearly 100,000 cells per cubic millimeter[1]. This immense neuronal density underscores the need for implantable neural recording technologies that can offer very high spatial resolution over different tissue spans.

Recently, ultra-high spatial resolution neural recording has been reported to improve the identification yield of cell types and enable recording from subcellular features such as axons and dendrites[2], but from local tissue spans. These findings suggest that combining higher spatial resolution with massively parallel recording across smaller, denser electrodes could yield a more comprehensive map of neural activity. Despite decades of miniaturization efforts and channel scaling in implantable neural recording technologies, the underlying design paradigms remain largely unchanged, constraining the number of simultaneously active recording channels.

A typical passive probe employs multiple metal electrodes along a shank, with each electrode requiring a dedicated interconnect wire to carry signals to external electronics. State-of-the-art passive probes, with shank widths of approximately 50 µm, can accommodate up to 64 electrodes[3]. To achieve high-density channels while keeping the cross section small, probes with wires as narrow as 300 nm have been demonstrated utilizing advanced microfabrication techniques[4]. But further miniaturization poses challenges, including fabrication limitations, increased noise, and signal crosstalk. Thus, wiring or the interconnect density fundamentally constrains scalability in passive neural probe architectures.

Moreover, the recorded analog electrical signal is vulnerable to external interference. To make it transmittable through long distance without major attenuation, the analog signal must be converted to a digital signal before transmitting, so the backend with analog-to-digital converter (ADC) is necessary to be attached to the probe. The size and weight of the backend put another challenge to the scalability considering the requirements from free-moving animals in behavioral experiments.

Active neural probes using complementary metal-oxide-semiconductor (CMOS) electronics with switchable electrode banks have been introduced, where only a subset of the available recording electrodes can be addressed at a time to provide more flexibility for large scale recording[5,6]. More recently, time-division multiplexing (TDM) based active CMOS probes have been reported, which can address all of the available recording channels[7]. TDM is a serial method of transmitting recorded signals one at a time, which suffers from noise-bandwidth tradeoff, limiting scalability. Moreover, active probe architectures with on-electrode amplification and on-shank multiplexers provide improved performance but those may come at the expense of high power consumption that can lead to tissue heating. These new architectures often rely on AC-coupled amplifiers, that unfortunately filters out ultra-low-frequency neural signals, such as infra-slow activity (ISA; <0.1 Hz). ISA has its unique role in brain function and is clinically significant [8,9] in conditions such as stroke, and migraine[10]. The capability to record ISA is crucial for effective diagnosis of different neurological disorders[11]. However, due to charge accumulation on the metal-electrolyte interface, leading to voltage drift, recording of DC shifts and ISA can be challenging with neural probes (either active or passive designs) with metal microelectrodes. Currently, recording of ISA and DC shift is limited to liquid-filled micropipette electrodes or Ag/AgCl electrodes, which can only offer a limited spatial mapping capability, and impractical for scalable long-term recording[12]. Recently, Graphene-based FET probes have demonstrated broadband DC-coupled recording, but their scalability remains subject to the same wiring and multiplexing limitations as conventional electrical probes.[8,12].

Together, these constraints highlight a fundamental problem: electrical probes are hitting physical and architectural limits that prevent dramatic scaling of the recording channels, even as neuroscience increasingly demands thousands or tens of thousands of simultaneous sites.

In contrast to electrical systems, optical communication routinely transmits enormous amounts of information over single waveguides by exploiting the ultrahigh bandwidth and multiplexing capacity of light. If sub-mV neural signals could be efficiently detected and directly transduced into the optical domain, one could leverage wavelength-division multiplexing (WDM) to route hundreds of independent channels through a single waveguide. Such a paradigm could enable massive scaling without the penalties of dense wiring or serial time-domain multiplexing. Optical transmission is also immune to electromagnetic interference, a persistent issue for electrical probes, where weak signals are highly vulnerable to ambient noise before amplification. Figure 1a and 1b show the envisioned system of of optically multiplexed neural probe, where information from many sensors is transmitted using light over only a few optical interconnects.

To realize this vision of an implantable optically multiplexed neural probe, we need to simultaneously achieve: (i) sub-millivolt sensitivity to resolve weak extracellular signals, (ii) DC-to-kHz bandwidth to capture both slow and fast neural dynamics, (iii) scalable optical multiplexing for large-scale recording, and (iv) compatibility with translation to an implantable form factor by using integrated photonic components for sensing and light in- and out-coupling.

To enable optical multiplexing for neural data transmission, the main critical step is to convert the electrical neural signal to an optical signal modulation. This is challenging and has never been done in an integrated platform, since the amplitudes of neural signals are very small- ranging from tens of micro-volts to a few millivolts. Therefore, extremely sensitive transduction mechanisms are needed to convert electrical signals into optical domain. A few research groups have recently looked into different mechanisms for this conversion. Optical fiber reflectometry has been theoretically proposed recently for the detection of neural signals[13]. Surface plasmon resonance (SPR) sensors have also been proposed to detect the small refractive index changes induced by the electrochemical changes due to neural activity near metal surfaces[14–17]. Graphene-integrated optical prisms have been used to optically detect sub-mV potentials from electrogenic cells cultured on their surface[18,19]. While these approaches demonstrate the promise of optics, they also face critical limitations. They mostly rely on bulky free-space optics components such as lenses, beam splitters, or prisms, therefore, making them incompatible for translation to compact, implantable probes. Moreover, these architectures are geared towards surface-level detection of neural signals, akin to surface electrocorticography, rather than penetrating interfaces needed for dense intracortical recording. Importantly, to date there has not been a demonstration of a unified platform that meets all four requirements for an implantable optically multiplexed neural probe.

In this work, we discuss an innovative solution to address all these issues through designing an ultra-sensitive electro-optics sensor that can directly detect and convert weak electrophysiological signals, from DC to high frequencies, to optical modulation. This novel electro-optic sensor provides the missing link that would enable realizing massively multiplexed neural recording probes that operate using a wavelength-domain multiplexing (WDM) scheme, where the signals from many sensors, each at a different wavelength of light, can be simultaneously routed through a single optical waveguide. Therefore, as opposed to either using many interconnect wires or serial time-domain multiplexing and interleaving of the signals (i.e., TDM), in this scheme, a single interconnect (i.e., an optical waveguide) can route the signals from many channels simultaneously. The optical multiplexing scheme mitigates the bottleneck of scalability from interconnect density and backend size, since a single optical waveguide can replace multiple electrical interconnects and a single optical fiber strand can transmit the optical signal with encoded neural signals over a long distance without attenuation.

The focus of this paper is to introduce a novel graphene-integrated electro-optic nanophotonic sensor capable of transducing sub-millivolt electrical signals into the optical domain. In this sensor architecture, the extracellular potential is captured by a graphene microelectrode situated on top of a photonic microresonator (Figure 1b, top inset). Then the electric potential across graphene is translated to a change of optical absorption of the monolayer graphene. The change of absorption in the graphene layer modulates the underlying photonic microresonator to encode information of the neural signals (e.g., action potentials) into the optical signal. Essentially, each of these microresonators operating at a particular resonance wavelength corresponds to a unique recording channel of the probe. A single optical waveguide (the bus waveguide) can be used to carry information from many of these microresonators, each with a slightly different size and a unique resonance wavelength. By utilizing WDM, we can passively multiplex information from multiple resonators via a single photonic waveguide (Figure 1b, bottom inset). Therefore, this novel design obviates the need for routing multiple interconnect wires to overcome the stringent requirement of 1-wire per 1-electrode in traditional passive neural probes.

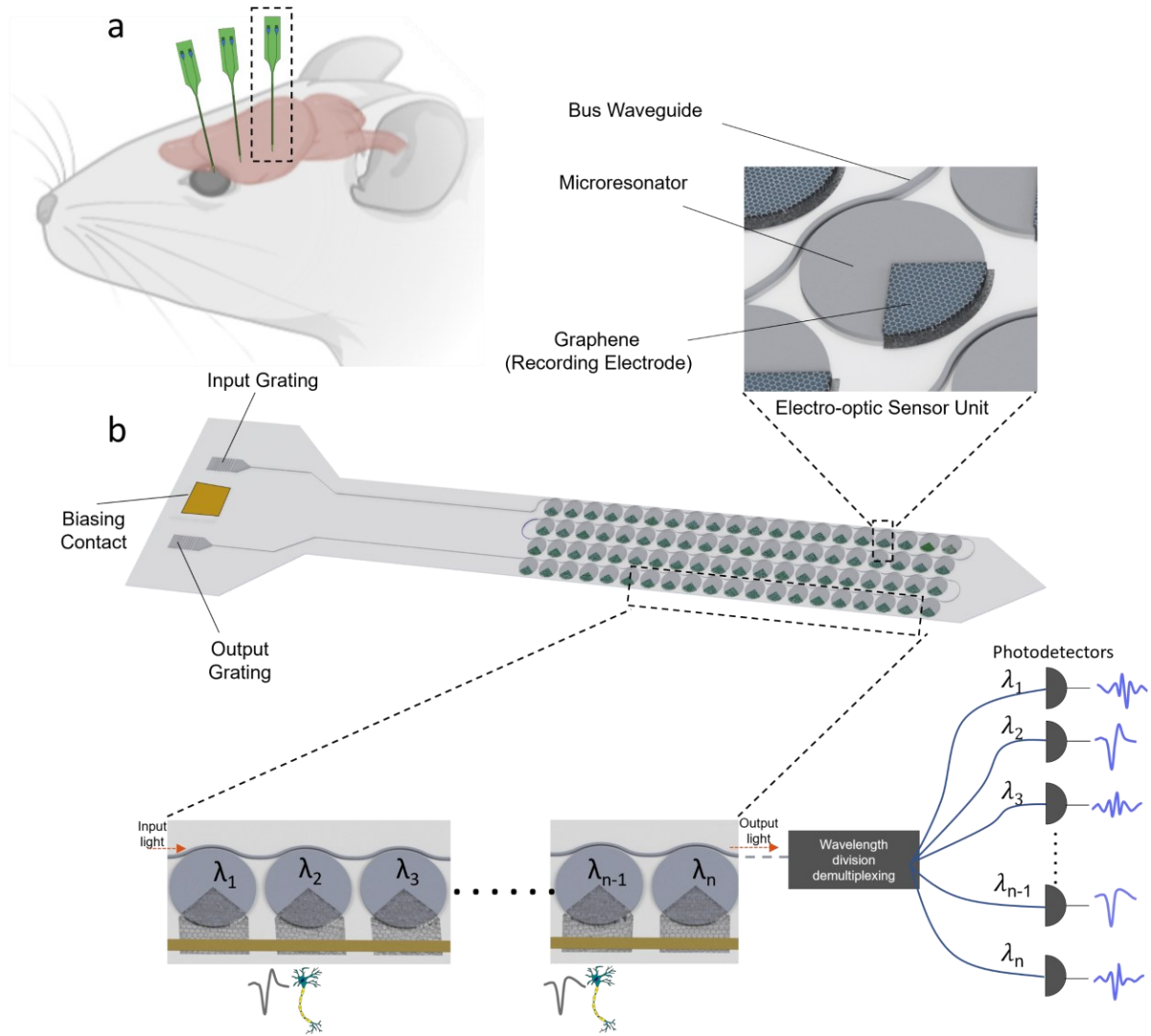

Figure *1* (a) Illustration of a rat implanted with multiple optically multiplexed neural probes. (b) Design schematic of the optically multiplexed neural probe, featuring an array of electro-optic sensors. Each sensor unit includes a graphene-integrated microresonator (top inset), where graphene functions as the recording electrode. Neural electrical signals are converted into optical signals at each sensor. Through wavelength division multiplexing (bottom inset), these optically converted signals from the sensor array are channeled through a single waveguide. At the waveguide's output, wavelength division demultiplexing is employed to reconstruct the electrical signals detected at each sensor.

Beyond its remarkable potential for scalability, WDM offers another unique advantage over electronic time-division multiplexing (TDM), which is immunity to electromagnetic interference. In active CMOS probes, electrical signals must be amplified and multiplexed by dedicated electronic circuits on or near the electrodes, leaving the weak analog signals highly susceptible to external noise. By contrast, our proposed probe architecture integrates signal transduction, amplification, and multiplexing directly at the recording site using electro-optic sensors (Figure 1b, top inset).

Neural signals are first converted from electrical to optical form by a monolayer graphene layer, then amplified by a microresonator through resonant light–matter interactions and finally encoded into wavelength channels. Because low-frequency neural activity is upconverted into optical modulation at hundreds of terahertz, the system achieves robust immunity to electromagnetic interference, including line-frequency coupling, a persistent challenge of recording with conventional electrical probes.

In this paper, we present a novel integrated photonic electro-optic sensor that demonstrates sub-millivolt sensitivity across a DC–kHz bandwidth, enabling the detection of both slow and fast neural dynamics. Each sensor consumes only 15 µW, achieving power efficiency comparable to state-of-the-art CMOS-based neural probes, such as Neuropixels[20,21]. Moreover, with advanced optical power sources and detection systems, these sensors can operate at even lower power levels, while maintaining high sensitivity. We validate the sensor performance through benchtop measurements and ex vivo brain slice recordings, showing that neural activity can be faithfully transduced into the optical domain. Beyond demonstrating the sensor unit cell, we also show multichannel optically-multiplexed neural recordings from mouse brain tissue to further demonstrate the feasibility of this architecture for scalable neural recording. Together, the results in this paper establish the foundation for a versatile, scalable neural recording platform that combines sensitivity, power efficiency, and scalability for implantable photonic neural probes.

## Results

### Design and operating principle of electro-optic neural sensor:

The unit cell of our neural sensor is a highly efficient electro-optic sensor consisting of a graphene-integrated photonic microresonator coupled to an optical waveguide (Figure 2(a)). The graphene-integrated photonic microresonator transduces electrophysiology signals into optical signals. When a nerve cell (i.e., a neuron) fires, the surrounding local electric field undergoes a rapid and dynamic change. The movement of ions across the neural membrane creates this local change in the electric field in the vicinity of nerve cells. Our novel electro-optic sensor leverages the unique properties of graphene to detect the small changes of electric field in the medium. The optical absorption of graphene continuously changes with electrostatic biasing; depending on the amplitude of the external electric field experienced by graphene, its Fermi level energy level is modulated, which in turn changes its optical absorption (Figure 2(c)). The Fermi energy level of graphene, $E_F$ is related to the applied electrostatic potential, $V_{bias}$ as

$$E_F \propto \sqrt{C_g(V_{bias} - V_{CNP})}, \qquad \text{Eq.(1)}$$

where $V_{CNP}$ is the charge neutral potential, and $C_g$ is the capacitance per unit area at the graphene-electrolyte interface. In our electro-optic sensor design, graphene serves as the recording electrode directly interfacing with the biological medium. As shown in Figure 2(b) the capacitance ($C_g$) is the electrical double layer (EDL) capacitance formed at the interface between the monolayer graphene and the physiological solution[22]. The capacitance of electrical double layer capacitor formed at the graphene electrode in a physiological medium is very high (> 2 µF/cm$^2$) due to the extremely small Debye length of ionic electrolyte[23,24]. This capacitance per unit area is much higher than that of conventional graphene based electro-optic modulators and field effect transistors utilizing dielectric materials as gate insulators. Due to this high capacitance in our electrolyte-gated graphene-integrated optical sensor, the Fermi level and optical absorption properties of graphene can be efficiently modulated in response to a small biasing potential. It can be seen from Eq. (1) that the larger the capacitance, $C_g$, the larger the change of Fermi energy level, $E_F$, in response to an input biasing potential. While the graphene's Fermi level can be efficiently tuned with electrostatic bias, the total change in optical absorption through a monolayer graphene is still small, and not sufficient for efficient electro-optic transduction of sub-milli-volt electrophysiological signals. To address this challenge, we integrated the monolayer graphene on a microdisc resonator. The photonic microresonator enhances the interaction of light with graphene by effectively increasing the interaction length as the light travels multiple times around the resonator. This enhancement is instrumental in addressing the issue of limited modulation of optical absorption in just a single graphene layer. Electrostatic biasing of optical absorption in graphene changes the round-trip optical loss in the photonic microresonator and also induces a phase shift. The change of round-trip optical loss causes a change in the intrinsic quality factor of the microresonator. Moreover, the induced phase shift results in a change of resonance wavelength of the microresonator. Both of these effects can be detected as a change of optical transmission (Figure 2(d)) through the coupled bus waveguide near the resonance wavelength of the microdisc resonator.

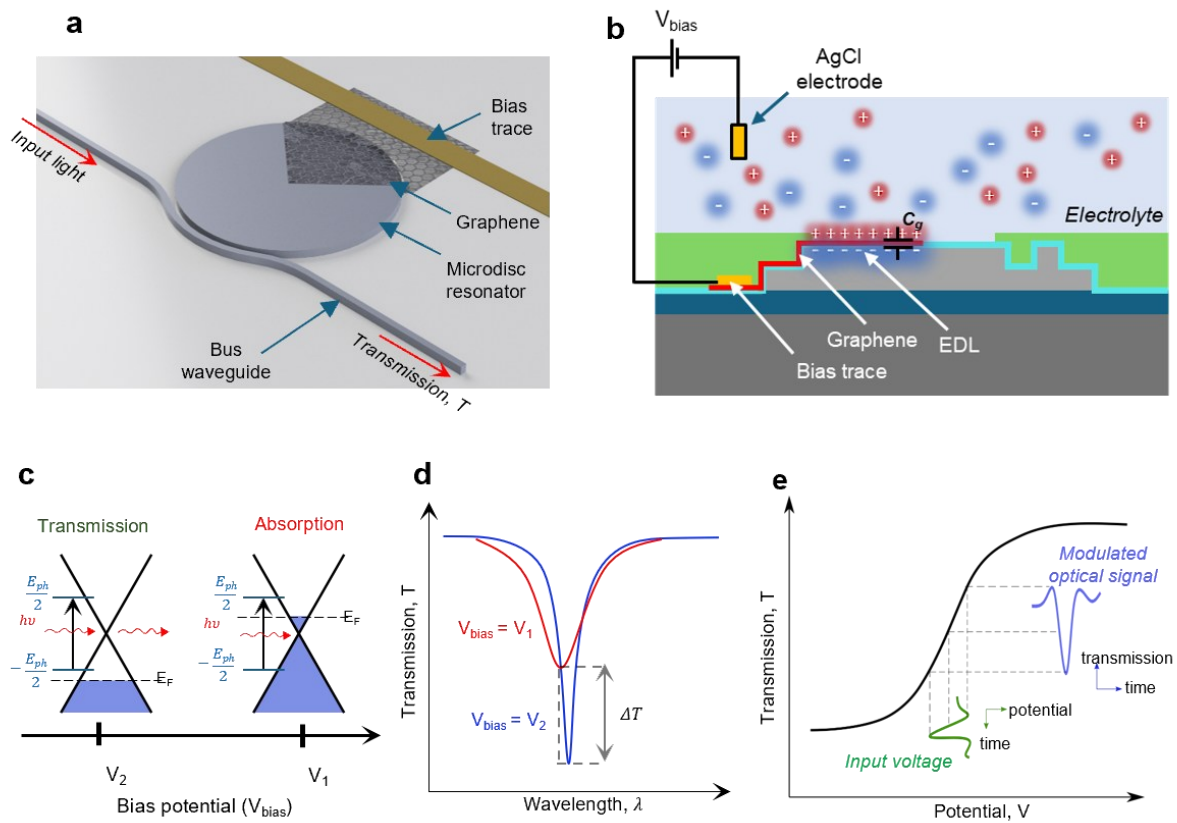

Figure 2 (a) Schematic of a unit cell of the electro-optic sensor. (b) A schematic showing electrical-double layer capacitance formed at the graphene-electrolyte interface as a bias voltage is applied for electrostatic biasing of graphene (c) Optical absorption modulation of graphene via Fermi level change at different electrostatic bias voltage. (d) Change of optical transmission at the bus waveguide coupled to the electro-optic sensor as $V_{bias}$ is changed. (e) A schematic diagram showing how small amplitude electrical signal at the electro-optic sensor gets transduced into optical signal via transmission modulation.

The transmission, T through the waveguide is related to the intrinsic quality factor, $Q_0$ and the resonance wavelength, $\lambda_0$, according to

$$T(\lambda, Q_0, Q_C) = \left| \frac{j2\lambda_0\left(\frac{1}{\lambda} - \frac{1}{\lambda_0}\right) + \frac{1}{Q_0} - \frac{1}{Q_C}}{j2\lambda_0\left(\frac{1}{\lambda} - \frac{1}{\lambda_0}\right) + \frac{1}{Q_0} + \frac{1}{Q_C}} \right|^2,$$  Eq. (2)

where, $\lambda_0$ is the resonance wavelength, and $Q_c$ is the coupling quality factor, which is determined by the interaction length and gap between the bus waveguide and the microresonator. As it can be seen from Eq. (2), the change in $Q_0$ and the resonance wavelength, $\lambda_0$, induced by the change in electric potential near graphene (through modulating its Fermi level) translates to a change in optical transmission. Figure 2(e) illustrates a conceptual transfer function of the electro-optic sensor, which shows how applied voltage to the sensor can be translated into a change in optical transmission intensity. A small-amplitude time varying input signal, like a neural signal, occurring in the vicinity of the electro-optic sensor would induce a proportional modulation of the optical transmission. The slope of this transfer function is the voltage sensitivity of the electro-optic sensor. The sensitivity can be enhanced by designing the whole sensor unit cell carefully to achieve an amplified optical response to the signal input. Here we discuss the approach to designing the electro-optic sensor to achieve a high level of sensitivity. The sensitivity of the electro-optic sensor can be defined as the ratio of relative change in the transmitted optical power through the bus waveguide (output) to the neural signal (input), which can be calculated as

$$\frac{\Delta T}{V_{neuron}} = \frac{dE_F}{dV_{bias}} \cdot \frac{dQ_0}{dE_F} \cdot \frac{dT}{dQ_0} \cdot T \cdot P_{in},$$  Eq. (3)

where, $V_{neuron}$ is the small amplitude time varying neural signal to be detected, $V_{bias}$ is the fixed biasing potential, $E_F$ is the Fermi Energy, level $Q_0$ is the intrinsic quality factor of the resonator, $T$ is the transmission and $P_{in}$ is the input laser power. The first term in the product relates the change of Fermi level to the biasing potential. As discussed earlier, the Fermi level is related to biasing potential ($V_{bias}$) via the capacitance per unit area of the electrical double layer capacitor formed at the graphene-electrolyte interface. Using graphene as the electrode surface for biological signal detection allows us to utilize the high EDL capacitance of the interface that contributes significantly to achieve the high voltage sensitivity in our electro-optic sensor architecture. The second term of sensitivity corresponds to $Q_0$ and $E_F$. The $Q_0$ of the graphene-integrated microresonator depends on the graphene coverage ratio, defined as the ratio of the area of the sector of the disc covering graphene to the surface area of the disc, as well as the electrostatic bias. Finally, $\frac{dT}{dQ_0}$ is a function of the coupling quality factor, $Q_c$, that is determined by the coupling gap between the bus waveguide and the resonator. If we normalize the sensitivity by the transmission $T$ and input power $P_{in}$, the sensitivity of the electro-optic sensor depends on careful choice of design parameters: graphene coverage ratio, biasing voltage, operating wavelength, and coupling gap.

Moreover, the voltage sensitivity is high in a narrow wavelength range around the resonance wavelength of the microresonator. Therefore, we can couple multiple electro-optic microdisc sensors, of different dimensions with different resonance wavelengths, to the same bus waveguide. As a result, the encoded electrical signals at each of the sensors will be spectrally separated at their corresponding resonance wavelengths. Consequently, the information from all these channels can be optically multiplexed and transmitted through the same waveguide. In other words, in addition to transducing the electrical signal into the optical domain, the electro-optic sensors also passively multiplex the encoded information into wavelength domain, to enable massively scaled recording of electrophysiological signals.

## Sensitive detection of sub-mV signals with electro-optic transducer

We have fabricated graphene-integrated electro-optic sensors on a 220 nm silicon-on-insulator (SOI) platform with a 2 μm buried oxide (BOX) layer. Figure 3(b) shows an optical microscope image of a sensor which consists of a 50 μm diameter microdisc resonator coupled to a 550 nm wide waveguide. The waveguides are terminated with grating couplers optimized for coupling transverse-electric field (TE) modes. A 7-nm layer of $SiO_2$ serves as a buffer layer encapsulating the photonic resonator, waveguide and gratings. Monolayer graphene is transferred onto the photonic resonator using a wet transfer process and then patterned to cover a sector of the microdisc resonator. In this figure, the graphene coverage is 25%. Part of the graphene structure extends beyond the microdisc resonator and connects to the common biasing metal trace. Finally, a 1.5 µm thick SU-8 layer covers the sensor and the metal trace, while only exposing the graphene-integrated microdisc resonator.

The voltage sensitivity of the electro-optic sensor is characterized in 1X PBS (phosphate buffer saline) solution, which is contained in a Polydimethylsiloxane (PDMS) chamber. Monolayer graphene is electrostatically biased through the PBS solution by applying an electric potential between a biasing pad and an AgCl pellet electrode placed in the electrolyte solution (Figure 3(a)). A tunable laser ranging from 1500 nm to 1630 nm provided optical input power to the sensor. The optical transmission is measured using an InGaAs photodetector and then amplified using an integrated electronic amplifier (Details in Methods section). Figure 3(c) shows the resonance spectra of the sensor as the monolayer graphene is electrostatically gated through the PBS solution using the AgCl pellet electrode. The bias voltage is increased from 150 mV to 600 mV in 50 mV steps. The optical transmission undergoes a dramatic change as the bias voltage is swept. An extinction ratio of ~ 7 dB is observed for a very small modulation voltage of ~550 mV. We also observe the shift of resonance wavelength with change of biasing potential. By probing the sensor over a range of wavelengths, with a small-amplitude sinusoidal input electrical signal at different biasing voltage amplitudes, we can extract the gain of the electro-optic sensor, $\frac{dT}{dV}$, and the SNR as a function of the wavelength and the bias voltage from these experimental results. As can be seen from Figure 3(d), the maximum gain of the sensor is obtained at a wavelength of 1589.68 nm and a bias of 600 mV. It appears that if we keep increasing the bias voltage around this wavelength, the gain will increase even further. However, since this wavelength is away from the resonance of the microresonator, the transmitted optical power is relatively high (Figure S3(a)). As a result, the relative intensity noise (RIN) of the laser is pronounced at this wavelength, as shown in Figure S3(b). This increased noise level results in a poor SNR, as shown in Figure 3(e), thus compromising the ability of the sensor to effectively detect neural signals. Therefore, there is a trade-off between optimizing for gain and reducing noise when selecting the bias voltage and operation wavelength. The SNR is maximized when the sensor is operated at 1589.59 nm, which is much closer to the resonance wavelength of 1589.57 nm at the bias of 450 mV. At this point, with lower transmitted optical power, the laser's intensity noise is also reduced, while the sensor still maintains a relatively high gain. Therefore, at this point, a balance of moderate gain and lower RIN results in the highest SNR.

We set the operating wavelength of the electro-optic sensor at this wavelength and bias it for optimum SNR. At this operational bias and wavelength, other performance metrics, such as the limit of detection and bandwidth are measured. To assess the limit of detection, we applied a small-amplitude, time-varying voltage (modulating signal) added to the optimum bias voltage. Both sinusoidal and square wave signals with varying amplitudes were tested. Our sensor successfully detected signals with an amplitude of 1 mV, achieving a high signal-to-noise ratio (SNR) of 30 dB (see Supplementary Figure S4). As shown in Figure 3(f), the root mean square (RMS) of the input-referred noise in 1x PBS is measured to be 26.13 ± 7.21 µV in the local field potential (LFP) band (1 Hz to 300 Hz) and 14.03± 5.55 µV in the action potential (AP) band (0.5 kHz to 7.5 kHz). Figure 3(g)shows the optical detections of sub-millivolt signals using our electro-optic sensor. Achieving an even lower detection limit is conceivable through the reduction and mitigation of diverse noise sources, including but not limited to laser wavelength jitter noise, laser intensity noise, photodetector noise, and ambient mechanical vibrations. Strategies to reduce noise will be elaborated in the discussion section.

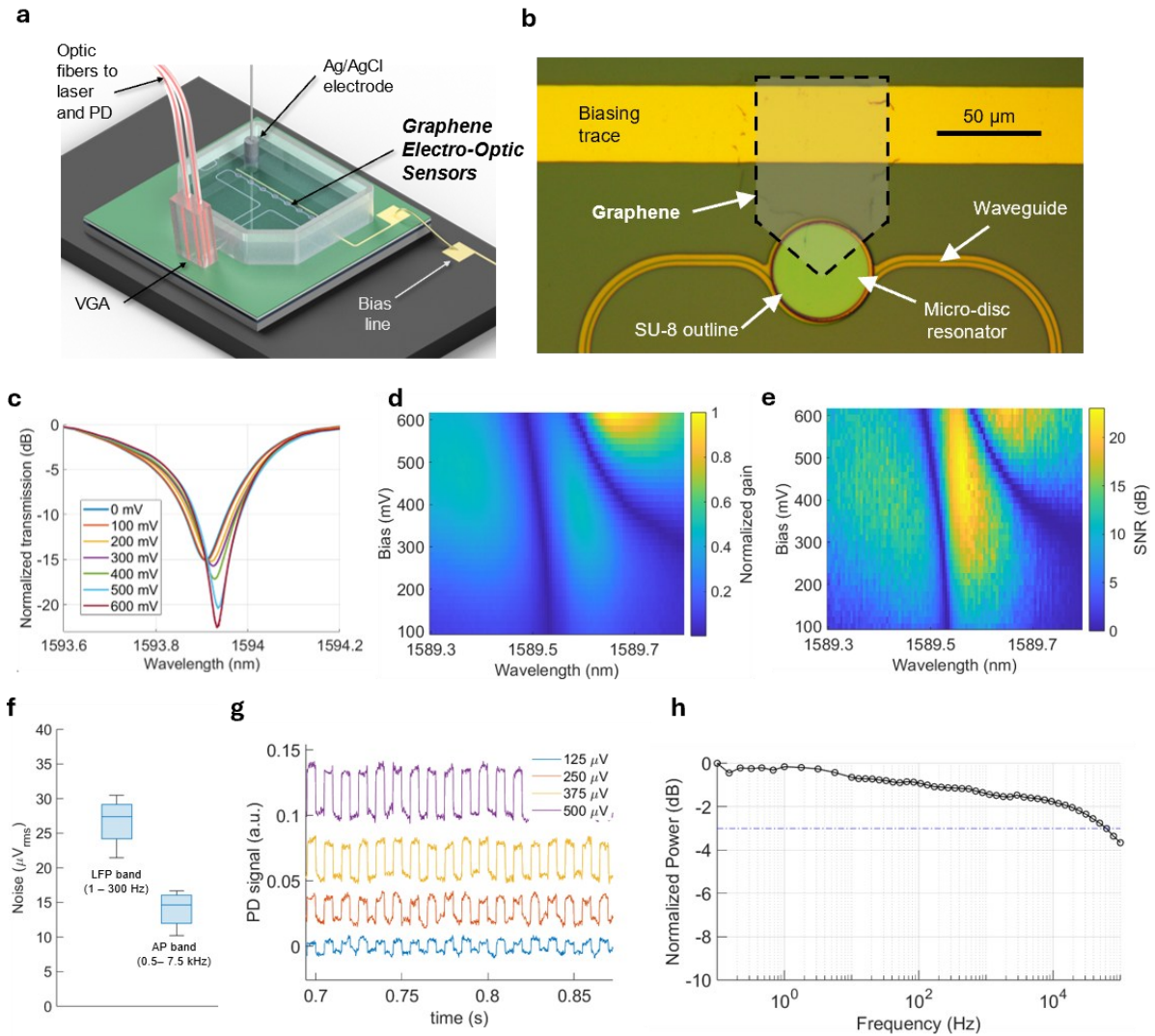

*Figure 3 (a) A schematic diagram illustrating the application of biasing potential and small signal modulation voltage for the electro-optic sensor's characterization. (b) Microscope image of the unit cell of the novel-electro optic sensor. (c) Graph showing the experimentally measured variation in the optical spectra of the electro-optic sensor with changes in biasing voltage. Surface plot showing experimentally measured (d) normalized small signal gain and (e) signal to noise ratio (SNR) as a function of the wavelength and bias voltage. (f) Detection noise(input-referred) floor of the electro-optic sensor in the LFP band and AP band of neural signal. (g) Recorded optical response from the electro-optic sensor to sub-millivolt modulating signals (h) Frequency response of the electro-optic sensor.*

The electro-optic sensor has a sufficiently high bandwidth for detecting both local field potentials (LFP) and action potentials (AP). The bandwidth of the sensor is mainly determined by the electrical double layer capacitance formed at the graphene-electrolyte interface and the resistivity of the monolayer graphene film. The surface area of the graphene exposed to the electrolyte is small. Therefore, despite having a large per unit area capacitance at the graphene-electrolyte interface, the total capacitance is rather small. As a result, the detection bandwidth of the sensor can be high, on the order of tens to hundreds of kHz, far exceeding what is needed for accurate measurements of neural signals. We have measured the frequency response of the electro-optic sensor by applying a 10 mV$_{pp}$ signal in the frequency range of 0.01 Hz to 100 kHz. We observed a mostly uniform magnitude response across the spectrum, as shown in Figure 3(h). Consistent magnitude response at lower end of the frequency (around 0.1 Hz) indicates the sensor's capability for detecting infraslow neural activity. Moreover, a 3-dB higher frequency cutoff at ~40 kHz highlights the sensor's ability to faithfully capture even the most rapid temporal dynamics of neural activity. Typically, commercial neural amplifiers used for detection of neural signals with passive neural probes have a high frequency cutoff at 15- 20 kHz and a low-frequency cutoff at 0.1 Hz. Therefore, our graphene-integrated electro-optic sensor boasts a broadband detection capability to accurately capture neural signals spanning from slowly varying DC shifts and infraslow activity up to fast temporal dynamics of extracellular neural spiking activity.

## Detection of neuronal response in brain slice via electro-optic transduction

Ion exchange across the neuronal membrane during activity induces a gradient change of local electric field potential that can be detected by the graphene electro-optic sensor. We have demonstrated that our electro-optic sensor can successfully record signals with a sub-millivolt detection limit and high enough bandwidth to capture the temporal dynamics of neural signals very well. The ex-vivo experiments were carried out by placing a mouse brain slice on the electro-optic sensor in a custom-designed chamber made of polydimethylsiloxane (PDMS) to perfuse ACSF (artificial cerebro-spinal fluid) as shown in Figure 4(a). Neurobiology experiments were carried out in 2 different neural networks. First, we tested the sensor in somatosensory cortex, which is a complex network where axons and somas are staggered and contain diverse cell types. To evoke neuronal response, a biphasic electric current was applied to layer 5 of the somatosensory cortex using a Pt-Ir bipolar electrode. The electro-optic sensor was positioned underneath the layer 2/3 region to record the downstream evoked activity. The axons from pyramidal cells in layer 5 extend to layer 2/3 and branch out to other superficial layers. To perform a control experiment, a tungsten monopolar electrode was placed above the layer 2/3 region (Figure 4(c)-(e)) to record the ground truth electrical neural signal. This electrical probe is connected to a conventional neural amplifier (RHD 512-channel recording controller, Intan Technologies). Biphasic current train (Figure S1(a)) with a pulse width of 0.2 ms and amplitude of 400 µA was applied at 0.2 Hz to get the optimal evoked excitatory postsynaptic potential (EPSP) signals [25]. Both the stimulation artifact and neural response could be identified from individual trials, as shown in Figure 4(f). Following the stimulation artifact of charging and discharging (Figure S1(b)) induced by the stimulation pulses, a pronounced evoked neural response in the cortical layer 2/3 can be detected by the electro-optic sensor, that matched very well with the ground-truth simultaneous electrical recording from the same region using the tungsten electrode. Figure 4(g) shows the side-by-side comparison of the single trial optically detected signal, measured by the electro-optic sensor, and the electrically detected signal obtained using the tungsten electrode. We have also conducted multiple stimulation trials to demonstrate the statistical significance of the electro-optic sensor measurements in comparison with the tungsten electrode recordings. Figure 4(i) shows averaged neural response from 12 stimulation trials in cortical brain slices, which are obtained by removing the stimulation artifact from the raw recording signals using a template subtraction method. Specifically, we extracted the characteristic waveform of the artifact by fitting a double-exponential function to its decay phase, clustering them to separate different decay templates when presenting, and averaging across the recordings. This template was then subtracted both from the optically-recorded and electrically-recorded signals, effectively eliminating the stimulation-induced transient while preserving the underlying neural activity. Figure 4(h) shows the corresponding subtracted templates, revealing that optically recorded signals stabilize significantly faster than electrically recorded ones. The decay time constant, defined as the time required to reach 10% of the maximum artifact amplitude, was measured to be 0.52 ms for optical recording and 14.17 ms for electrical recording. This indicates that the transient artifact decays much faster in optical recordings, allowing for rapid extraction of neural signals immediately after stimulation. The faster stabilization of optically recorded signals highlights another unique advantage of the presented neural recording paradigm that leverages the wide frequency bandwidth of graphene-based electro-optic sensors in minimizing the effect of stimulation artifacts and improving the temporal resolution of neural activity recordings

The ex-vivo recording capability of the electro-optic sensor was also tested using a well-aligned neural network, i.e., the Schaffer collateral pathway in hippocampus. This pathway consists of axons from pyramidal cells that connect CA3 region to CA1. A biphasic electrical current stimulation (with the same parameters used for cortical stimulation) was applied to CA3 using a platinum-iridium (Pt-Ir) bipolar electrode, while the electro-optic sensor was positioned beneath the CA1 region. As a control, a tungsten monopolar electrode was placed above the CA1 region, near the electro-optic sensor, as illustrated in Figure 4(j) and (k).

The electrical currents stimulated the output axons of CA3 neurons, causing action potentials to propagate through the Schaffer collaterals and evoke neural responses in the CA1 region[26]. Biphasic current stimulation was delivered over 60 trials, with neural signals recorded using both the electro-optic sensor and the tungsten monopolar electrode. Figure 4(i) shows the subtracted templates of both recorded signals, demonstrating a shorter recovery time for optical recordings, similar to cortical recordings. Notably, the optical recording in the hippocampus exhibited a faster decay rate, with a time constant of 8.64 ms compared to 41.77 ms for the electrical recording, indicating more rapid suppression of stimulation artifacts. Figure 4(m) presents the average neural responses from 60 stimulation trials, recorded with both the electro-optic sensor and the nearby tungsten monopolar electrode. The evoked response, observed as an EPSP, appeared as a negative peak with a 3 ms latency following the stimulation pulse, indicating the depolarization of CA1 pyramidal neurons in response to CA3 stimulation. Results showed that the electro-optic sensor successfully detected the neural response, evident from the negative peak following the stimulation artifact, and demonstrated strong correlation with the recordings from the tungsten electrode. Minor variations, such as differences in magnitude and delay are attributed to differences in sensor placement and recording depth.

To verify the validity of the recorded neural responses, a control experiment was performed using a non-functional brain slice kept in a non-perfused saline solution until it no longer showed any evoked neural response detectable by the tungsten electrode. The electro-optic sensor also did not record any neural response. Furthermore, when the current was applied with reversed polarity, no neural activity was observed, and only the polarity of the artifact was reversed (Figure S 2). These control experiments, along with the hippocampal and cortical recordings, validate the electro-optic sensor's ability to consistently and accurately capture electrophysiological signals.

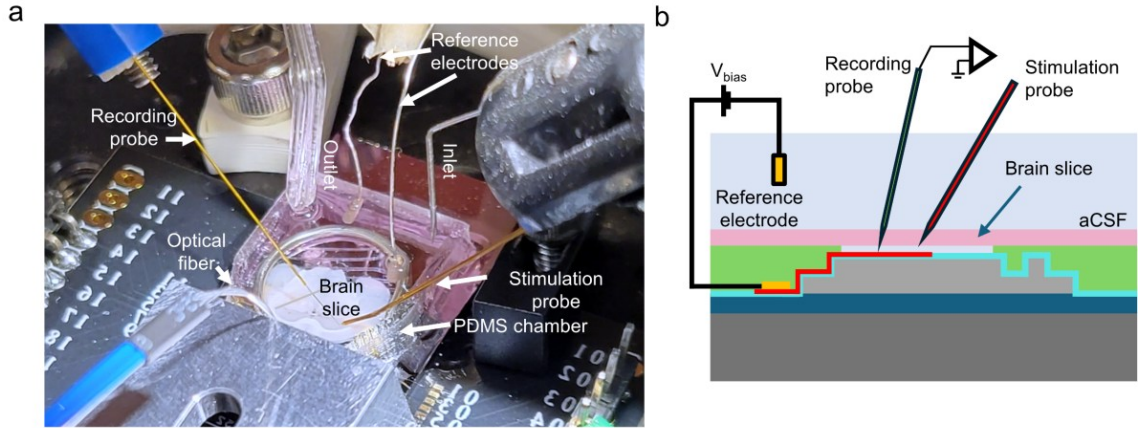

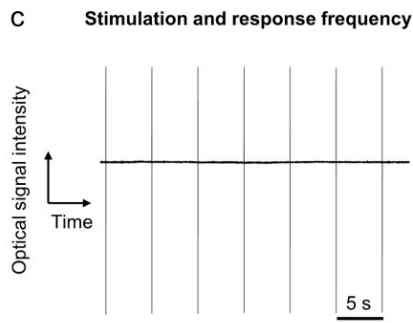
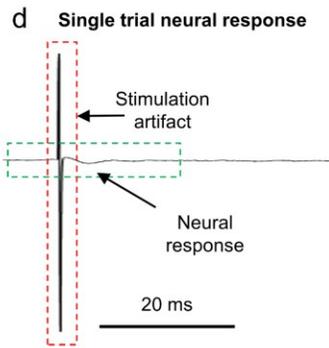
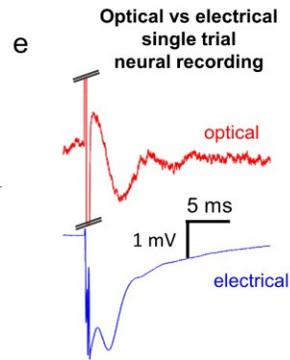

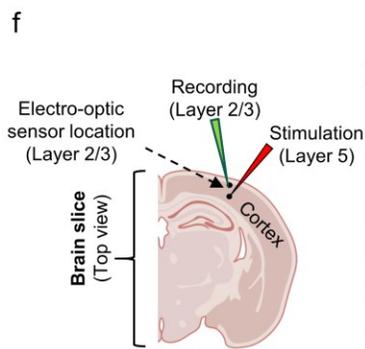
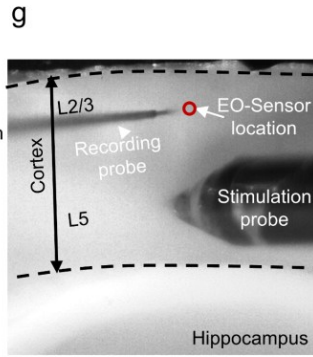
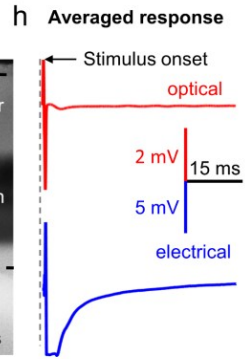
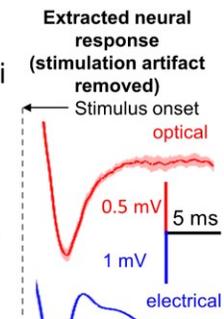

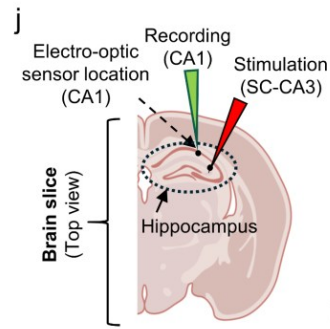
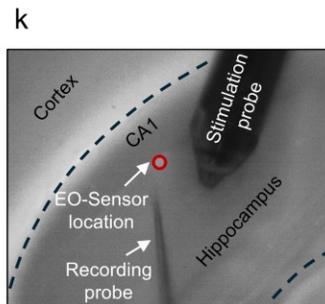
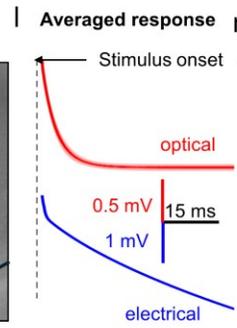
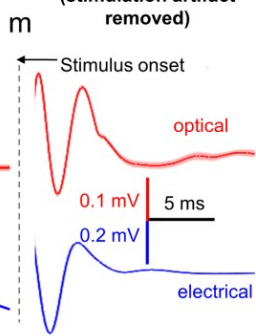

*Figure 4 (a) Experimental configuration for ex vivo recording from a mouse brain slice using the advanced electro-optic sensor. (b) Schematic cross section of the electro optic sensor showing the placement of stimulation and recording electrodes on brain slice. (c) Captured optical signal from the electro-optic sensor in the mouse brain slice during biphasic current stimulation at 5-second intervals. (d) An enlarged view of a segment of the recorded signal (single trial) is provided, where (e) illustrates the optically recorded (red) neural response that occurs approximately 5 ms following the stimulus onset. Concurrently measured neural signal from a Tungsten electrode (blue) placed at layers 2/3 of cortex. (f) and (g) showing diagram and microscope image depicting the positioning of the brain slice over the electro-optic sensor, with the stimulation electrode delivering biphasic current positioned in the cortex's layer 5, and the electro-optic sensor and tungsten recording electrodes situated in layers 2/3. (h) Averaged response (N=12) consists of stimulation artifacts and neural signals from optical (red) and electrical (blue) recordings. The optical recording shows significantly faster artifact suppression, with a decay time constant (time to 10% of max) of 0.52 ms versus 14.17 ms for electrical recording. (i) Trial averaged response (N=12) of optically measured (red) and electrically recorded (blue) neural signals showing good correlation. The neural signals are extracted after stimulation artifact is removed in post-processing utilizing a template subtraction method. The template for stimulation artifact is constructed by averaging multiple trials of raw neural signals captured with optical sensor and electrical recording probe. (j) and (k) present a schematic and a microscope image, respectively, showing the arrangement of the stimulation electrode, tungsten recording electrode, and the electro-optic sensor on the hippocampal region of the slice. (l) Averaged (N = 60) response consists of stimulation artifacts and neural signals recorded in the hippocampus using optical (red) and electrical (blue) methods. The optical recording exhibits faster artifact suppression, with a decay time constant (time to 10% of max) of 8.64 ms compared to 41.77 ms for electrical recording. (m) displays the averaged response from the brain slice, measured from 60 trials, as captured by both the optical sensor (red) and the electrical recording probe (blue) after removing stimulation artifact by template subtraction.*

## Multiplexed recording of LFP propagation in brain slice

To evaluate the scalability of our design, we fabricated a 10-channel electro-optic sensor array and tested it in ex-vivo brain slice experiments (Figure 5(a)). The fabricated microresonators span a radius range centered at 25 µm, with successive increments of 5 nm, yielding spectrally distinct resonance wavelengths. Benchtop optical characterization confirmed that each microresonator exhibited a unique resonance within one free spectral range (FSR), allowing a one-to-one mapping between resonance wavelength and physical sensor channel (Figure 5(b) and (c)). A single shared bus waveguide was used for delivering light and also readout of these sensors based on a wavelength division multiplexing scheme. Optically detected neural signals from individual sensors were multiplexed through this common waveguide and then demultiplexed after detection, enabling simultaneous recordings from multiple sites.

The array was placed beneath the brain slice and aligned such that individual sensors captured responses from distinct regions of both the somatosensory cortex and hippocampus (Figure 5(d) and (e)). For stimulation, a biphasic current at 0.2 Hz was applied to the CA3/Schaffer collateral region of the hippocampus, with individual pulses of 0.2 ms width and 400 µA amplitude (Figure S1(a)) to reliably evoke excitatory postsynaptic potential (EPSP) signals in the hippocampal circuits. The electro-optic sensors detected robust stimulus-locked evoked responses in the Schaffer collateral and CA1 regions of the hippocampus (Figure 5(e)). As expected, no significant responses were observed in cortical layers, reflecting the absence of direct synaptic connectivity between the stimulation site and the cortical regions for the brain slice under test.

Figure 5(f) presents the averaged optically detected neural responses from 60 stimulation trials (after artifact removal) across different channels, along with the heatmap (Figure 5(g)) showing the spatial distribution of normalized response amplitudes. The highest neural response amplitudes were recorded from the sensor channels located in the Schaffer collateral layer, which is closest to the stimulation site, followed by progressively weaker responses in CA1. The cortical layers showed no significant response during CA3/Schaffer collateral stimulation.

These results highlight the ability of our electro-optic platform to capture spatially resolved neural activity across multiple brain regions using only a single optical waveguide. The demonstration of robust wavelength division multiplexed recording in a 10-channel array supports the feasibility of scaling this approach to much larger arrays utilizing individual waveguides shared among multiple sensors.

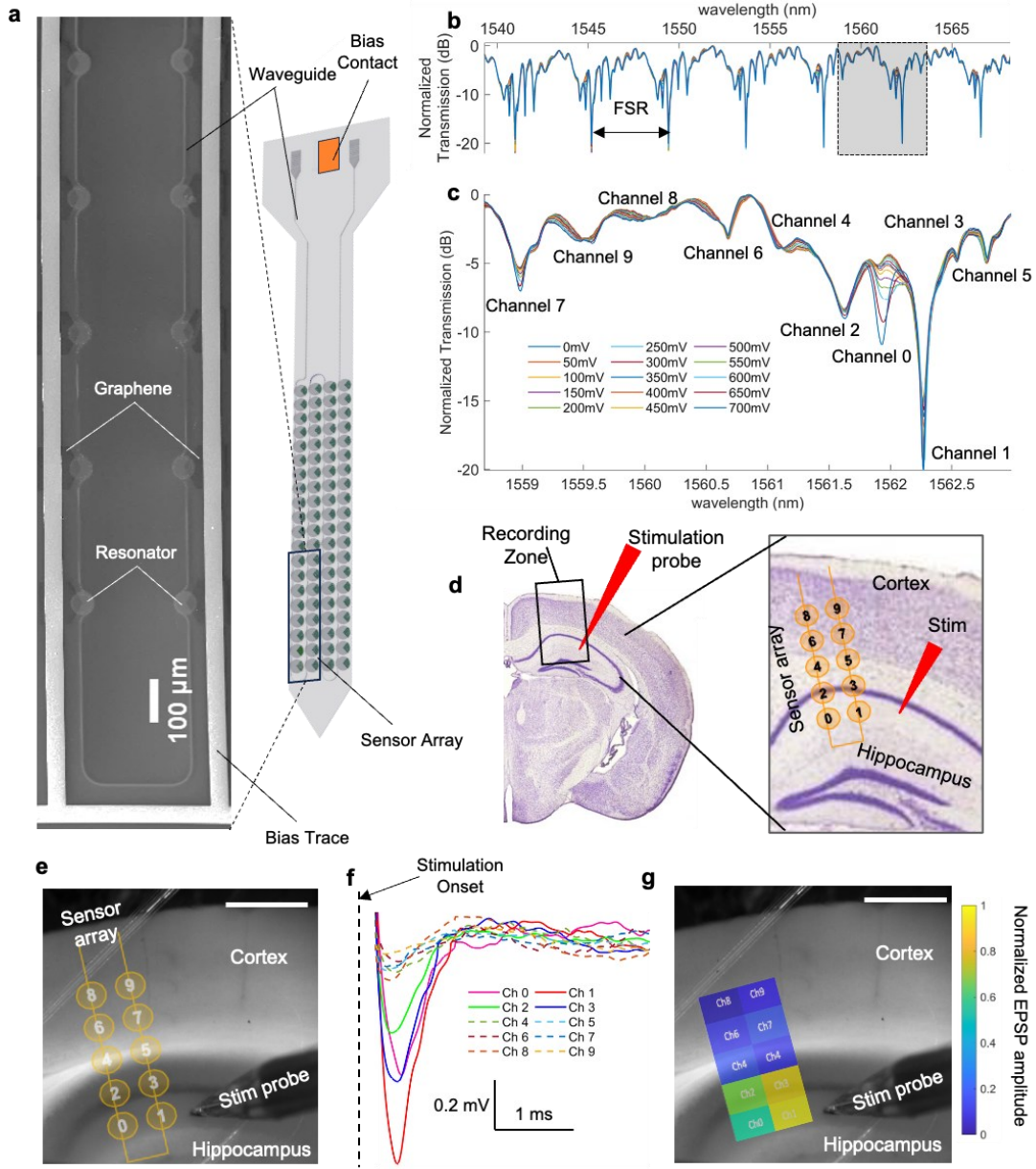

*Figure 5 (a) Scanning electron microscopy (SEM) image of the 10-channel electro-optic sensor array, showing ten graphene-integrated photonic microresonators coupled to a single bus waveguide. Each microresonator functions as an individual recording electrode and has distinct resonance frequency, enabling wavelength-division multiplexing of neural signals on one waveguide. (b) Benchtop characterization of the sensor array. Transmission spectrum of of the device spanning multiple free spectral range (FSR). (c) Ten distinct channels are identified within single FSR and mapped to physical locations. (d) Schematic and microscopic image (e) of the region of interest in brain slice positioned over the sensor array. A bipolar stimulation electrode was positioned in the CA3/Schaffer collateral region of the hippocampus. The sensor array was aligned to span both hippocampus and somatosensory cortex. The scale bar corresponds to 500 µm. (f) Optically multiplexed neural recording of evoked neural activity from the 10-channel sensor array. Averaged EPSP waveforms from 60 stimulation trials are shown after removing the stimulation artifact by template subtraction. Hippocampal channels are plotted as solid lines and cortical channels as dashed lines. (g) Spatial distribution of normalized EPSP amplitude response across different locations/channels. The scale bar in (e) and (g) correspond to 500 µm.*

## Discussion

Dense multiplexing can be realized by designing multiple electro-optic sensors, each with a slightly different microresonator diameter, such that they operate at different resonance wavelengths. This paper demonstrates the feasibility of this approach by integrating ultra-sensitive graphene microresonator sensors into a compact, scalable photonic platform. A single bus waveguide can transmit signals from many channels simultaneously, and we validated this concept by multiplexing ten wavelength-selective sensors based on a conservative design of ~50 µm-diameter resonators. Each sensor detected electrophysiological signals as small as ~25 µV (at 3 dB SNR) over a DC–kHz bandwidth, providing a clear proof-of-concept for the feasibility of massively parallel neural recording.

The maximum number of channels coupled to a common bus waveguide is determined by the resonator free spectral range (FSR) and its resonance linewidth, which is determined by its quality factor. The quality factor of the graphene integrated microresonator is dependent on several design factors, including the graphene coverage area over the microresonator and the thickness of the intervening $SiO_2$ buffer layer. Beyond these design parameters, resonator sidewall smoothness, residues on the surface of graphene, as well as the quality of the transferred monolayer graphene, also affect the loaded quality factor[27–29] and the performance of the sensor. Through meticulous tuning of these variables, it is possible to create microresonators with superior quality factors and reduced linewidths, allowing for multiplexing of more channels onto a single waveguide. Additionally, adopting resonators with a smaller diameter can be another effective strategy to increase the number of channels coupled to a single bus waveguide. For instance, an 8 µm-diameter microresonator has an FSR of about 30 nm. Our calculations suggest a loaded quality factor of ~$1.5 \times 10^4$ for a graphene-integrated microresonator (with optical loss mainly dominated by the absorption of graphene), leading to a linewidth of ~100 pm at the 1550 nm wavelength. This means that we can potentially have nearly 300 optically multiplexed recording channels over a single bus waveguide, while ensuring their resonance wavelengths are spectrally distinct and distributed over a single free spectral range. With just seven waveguides, up to 2100 independent recording channels can be recorded simultaneously on a probe shank with a footprint of 60 μm × 20 μm × 6.5 mm. Adding an additional waveguide (~500 nm wide) for 300 more channels would only add 1.5 μm to the probe shank width, while the microresonator takes up ~10μm. The microresonators can be designed to have smaller footprints to achieve even higher density of recording channels. However, it is noteworthy that as the resonator dimension is decreased, the intrinsic quality factor, $Q_0$ is reduced because the radiation and scattering losses increase. As a result, the sensitivity of the sensor can be compromised. Therefore, there is a trade-off between scalability of channels and the sensitivity and detection limit.

Scaling of the number of channels is indeed feasible based on prior demonstrations of dense photonic multiplexing; for example, on-chip microdonut spectrometer architectures have been demonstrated with more than 80 resonators coupled to a single on-chip optical waveguide[30]. However, dramatic scaling of the number of resonators in an array requires addressing some practical challenges since fabrication imperfections and wafer-scale thickness variations inevitably shift the resonance wavelengths across nominally identical elements. While thermal tuning is a typical remedy to adjust the resonances back[31,32], equipping each resonator with its own heater is power-intensive and spatially prohibitive, particularly for the implantable probe architecture discussed in this work. Fortunately, emerging post-fabrication trimming techniques offer efficient, low-overhead alternatives[33,34]. Localized laser annealing of cladding layers enables permanent wavelength adjustment with high precision and minimal added complexity[34]. Femtosecond laser trimming has also been used effectively to compensate for fabrication errors by locally modifying the refractive index of the resonator's cladding or structure, achieving resonance alignment within a free-spectral-range[33]. These techniques, combined with the rapid maturation of silicon photonic foundry processes, point to significantly improved process control over resonance uniformity, obviating the need for active thermal tuning. Our sensor design is inherently compatible with these industry-standard fabrication processes, requiring only minimal post-foundry steps to integrate graphene. Moreover, wafer-scale graphene integration into silicon photonics platforms has been realized in CMOS pilot lines, with hundreds of electro-absorption modulators demonstrating high yield reproducibility in a 300 mm pilot process[35]. This demonstrates the manufacturability of graphene-on-SOI devices at scale and fosters the promise of our proof-of-concept demonstration for translation into high-volume production with massive optical multiplexing.

Another important design consideration for massive multiplexing of neural recording using the electro-optic sensor is the optical input and output detection scheme. In our proof-of-concept experiments, we employed a tunable laser to select a specific wavelength, along with a photodetector to capture the optical signal at this wavelength. However, for the simultaneous recording of hundreds or thousands of multiplexed channels, the optical excitation and detection system needs to be designed differently. An optical input design could involve using an array of single-tone laser lines, with each laser line precisely tuned for a specific microresonator. For example, an array of tunable vertical cavity emitting laser diodes (VCSELs) connected to the bus waveguide through a waveguide-based optical power combiner[36] can be used as the optical input to the sensor array. Alternatively, a dense frequency comb generator can be used as another method of providing input to the electro-optic sensor array[37,38]. For the output, a dense wavelength demultiplexer (DWDM) can be used to segregate and direct light at different wavelengths corresponding to different channels to individual detectors. This approach enables simultaneous and parallel recording of all channels. The photodetector signal of each channel can be sampled at a typical sampling rate of 30 kSamples/sec. Another viable method involves using a swept-source laser capable of rapidly scanning the relevant wavelength range at a high repetition rate (2 kHz – 400 kHz)[39,40], a technique commonly utilized in optical coherence tomography[41]. This allows for swift scanning of all resonances at a high sampling rate. Concurrently, at the output end of the bus waveguide, a high-bandwidth photodetector can be used to record the temporal evolution of the entire spectrum. We can decode the neural signals from each channel by analyzing the modulation of their respective resonance line shapes over time. This approach significantly streamlines the input and output setup, requiring only one swept source laser for the input optical power and just one photodetector per bus waveguide. However, one challenge of using this approach is the extremely high bandwidth requirement for detection, which increases the noise floor of the measurement. This is because, in a swept source laser, the wavelength is tuned at microsecond time scales, which necessitates a measurement bandwidth of few hundreds of MHz. In addition to the increased noise from the large measurement bandwidth, there are additional noise contributions from phase noise, which stems from wavelength inaccuracy from one wavelength sweep to the next. Therefore, additional hardware or post-processing calibration methods need to be applied to compensate for such noises[42–44].

The noise floor of the optically transduced neural signal in the electro-optic sensor is mostly determined by the noise from the optical sources (i.e., laser sources) and detectors (i.e., photodetectors, data acquisition system). The optical source noise mainly constitutes the intensity and phase noise of laser source. The neural signal is transduced from electrical domain to optical domain by changing the optical transmission of the microresonator coupled to the bus waveguide. As we discussed earlier in the paper, the transmission change can arise from the intensity modulation and the wavelength modulation. Depending on the quality factor of the microresonator and the graphene coverage ratio, one of these two effects can be pronounced and therefore, either the laser intensity noise or the phase noise could become the dominant source of input noise. Besides the noise from optical input source, the noise of the photodetection system is also important. The noise-equivalent power (NEP) of the photodetector is the minimum required input power for the SNR to be unity at the photodetector output, which represents the limit of detection

(LOD) of the system at the output. Here we only consider the noise level in the corresponding frequency band, which is the local field potential (LFP) band (1 Hz to 300 Hz) and the action potential (AP) band (0.5 kHz to 7.5 kHz). Since we can use bandpass filter to remove noise from unrelated frequency bands in the postprocessing, only the noise in the same frequency band would directly affect the LOD. Utilizing a photodetector with a lower NEP, characterized by reduced shot noise, dark current noise, and resistive thermal noise, can significantly enhance the measurement accuracy of our system. Therefore, combining a laser with minimal intensity and phase noise with a low NEP photodetector can effectively lower the system's noise floor and improve its SNR. If the system noise is primarily dominated by the photodetector, increasing the optical power input will enhance the SNR, since the sensor sensitivity is directly proportional to the transmitted optical power. However, this approach results in higher optical power consumption by the electro-optic sensor. In our experiments, we have achieved a LOD of 14.03 µV for SNR =1 in the action potential (AP) band, with the corresponding optical power consumption of only 15 µW by using a detector with an NEP of approximately $0.2 \times 10^{-12}$ W/(Hz)$^{0.5}$ and tuning the laser at the operating wavelength with 0.1% optical intensity noise, which consists of relative intensity noise (RIN), phase noise and environmental noise such as mechanical vibrations coupled into the laser. The electronic noise and quantization noise from the data acquisition are negligible compared to optical intensity noise according to instrument specifications and our experiments. Optical intensity noise is the dominant source of noise in our measurements, contributing to nearly 90% of the overall noise. Power consumption per channel, therefore, can be reduced by utilizing a laser with lower intensity noise and phase noise to significantly reduce the power consumption per channel while retaining the LOD. A laser with 0.05% intensity noise can reduce the required laser power and the power consumption per channel to less than 11 µW without losing LOD. This reduction is crucial for allowing multiplexing and simultaneous recording from a large number of channels, as it avoids inducing tissue heating near the probe.

Overall, this work demonstrates an innovative electro-optic sensor platform capable of DC-to-kHz neural recording with sub-millivolt sensitivity, low power consumption, and multiplexing via WDM. The performance was experimentally demonstrated using ex-vivo experiments on mouse brain slices, accurately recording neural signals comparable to ground truth electrophysiology recordings. By combining high sensitivity, scalable architecture, and foundry-compatible fabrication with graphene integration, this approach provides a promising path for massively parallel, minimally invasive neural recording. Our findings highlight the potential of this WDM-based neural probe architecture in advancing neural recording technologies, capturing the complex neural dynamics of the brain with unprecedented spatio-temporal resolution. The tag-free nature of the electro-optic neural sensing, coupled with the miniaturization offered through WDM that can significantly reduce the implant footprint and tissue damage, makes this method potentially translatable to clinical applications.

# Acknowledgement

This material is based upon work supported in part by the National Science Foundation under Grant No. 1926804. The authors acknowledge the use of the Bertucci Nanotechnology Laboratory at Carnegie Mellon University supported by grant BNL-78657879.

# Disclosure

The authors declare no competing interests.

## Methods

### Fabrication of electro-optic sensors:

Graphene based electro-optic neural sensors are fabricated on a Silicon-on-insulator (SOI) chip (2 cm × 1.5 cm) with 220 nm device layer and 2 µm buried oxide layer. High resolution electron-beam lithography is used to pattern the photonic devices consisting of waveguides, gratings, and resonators. With an optimized Inductively Coupled Plasma (ICP) silicon etching process, the photonic devices are realized by ~100 nm deep shallow etching in silicon. Using partial etch instead of full silicon etch for photonic devices reduces scattering loss to improve quality factor and minimizes step height to achieve higher yield in graphene transfer and reduces likelihood of graphene tearing. Silicon out of the photonic devices is fully etched in an additional ICP silicon etching process with 3 µm ridge width. Then, a thin 7-nm layer of $SiO_2$ is deposited via atomic layer deposition (ALD) process as the buffer layer. Afterwards, a monolayer graphene film is transferred onto photonic devices using the conventional wet transfer process. The details of the process can be found on [45]. Once the transfer process is completed, graphene is removed from unwanted regions with photolithographic patterning and oxygen plasma based etching processes. Then, Ni-Au (5 nm- 150 nm) is evaporated and patterned with lift-off process to realize the biasing interconnect and pad on graphene. Finally, a 1.5 µm thick SU-8 layer is spin coated and patterned to insulate the photonic devices and the biasing traces while selectively exposing and defining the graphene microelectrode opening. Figure S5 shows a simplified process flow for fabrication of optically multiplexed neural probe.

### Optical Characterization:

The voltage sensitivity of the electro-optic sensor is characterized in 1X PBS (phosphate buffer saline) solution which is contained in a 3D printed PDMS chamber. The layout ensures that the graphene integrated resonators are exposedto PBS solution, while ensuring that the optical gratings remain isolated from the solution to facilitate efficient optical coupling from the fiber-coupled laser. Monolayer graphene is electrostatically biased by applying an electric potential between a biasing pad and AgCl pellet electrode placed in the electrolyte solution. The biasing potential is supplied by a function generator (33500B, Keysight). Light is coupled in and out of the input and output gratings respectively using a V-Groove Array (VGA) to measure the optical response of the waveguide coupled microresonators. A tunable laser (1500 nm- 1630 nm) (TSL-510, Santec) is used for exciting the sensor while an amplified InGaAs photodetector (2011-FC, Newport) captures optical power transmitted through the resonator coupled bus waveguide. The optical power is recorded by a DAQ board (PCIe-6374 and BNC-2110, National Instruments) with a sampling rate of 100 kSamples/second.

## Biological validation of electro-optical sensing system

To demonstrate the functionality of electro-optical sensor in biological tissue, an ex-vivo brain slice model was chosen which enables the controlled environment, qualifies precise examination of neural activity and excitability. For this, 3 months old C57BL/6 mice were used and experiments were performed in accordance with the Institutional Animal Care and Use Committee guidelines. For ex-vivo brain slice preparation, animals were anesthetized using isoflurane and decapitated. Further brain was isolated and bathed in ice cold aCSF (artificial cerbrospinal fluid). Brain slices of 350um were prepared using Vibratome (Leica V1200S) and transferred to PDMS recording chamber with the electro-optic sensor after 45 minutes incubation at room temperature in ACSF.

### *Stimulation:*

A special recording chamber with 18 mm length and 16 mm width was fabricated to study neural activity in brain slices with opto-electric sensor as shown in Figure 4 (a). To demonstrate efficacy of opto-electric sensor neural response was recorded in two different brain regions i.e., cortex and hippocampus in brain slices. Electrical stimulation was delivered using a Pt-Ir electrode to evoke neural response which furthered measured using electrical and optical sensor as extracellular field potential i.e., local field potential (LFP) For cortical recordings, stimulation was delivered in Layer 5 and evoked response was recorded in Layer 2/3 of somatosensory cortex whereas for hippocampal recordings, stimulation was delivered in CA3 regions and electrical and optical response was recorded in CA1 region using tungsten metal electrode and optical sensor respectively. For stimulation, a biphasic current pulse (0.2 ms pulse width, 0.2 Hz) with varying amplitudes were applied and for electrical response tungsten metal electrode was connected to a neural amplifier headstage (RHD 32-channel headstage, Intan Technologies). The optical responses from the electro-optic sensor were obtained by fixing the wavelength of laser near the resonance wavelength. A template subtraction to fit the rising and falling part with exponential functions is necessary for both electrical and optical responses to show the evoked local field potentials (LFPs).

### *Biological preparation and measurements:*

The animal model chosen for this study is the C57BL/6 (wildtype) animal. Mice used for the experiments were in accordance with rules of IACUC. Ex vivo experiments were chosen to allow for a controlled environment, enabling precise examination of neural activity and excitability. C57 wildtype animals underwent anesthesia with Isoflurane to ensure humane and controlled conditions for subsequent procedures. The brain was then dissected, and extraneous regions were carefully removed to focus on specific brain areas relevant to the study. The Vibratome, a precision cutting instrument, was employed to create brain slices with a thickness of 350 microns. The slicing process was performed at a speed of 20 mm/s to maintain consistency across samples.

The dissected brain, now reduced to essential regions, was securely affixed to a stage using superglue and placed in an ice cold oxygenated artificial cerebrospinal fluid (ACSF) chamber providing an environment conducive to the viability of neural tissue. The reason behind using ice cold ACSF is to slow down the metabolic process of the tissues, inversely helping in less damage to the cells. The slices are then incubated at room temperature for 45 min to an hour to recover neural activity to normal before the experiment[46,47].

To evoke the extracellular field potential, biphasic current pulses (0.2 ms pulse width, 0.2 Hz) with varying amplitudes were applied to the CA3 region of hippocampal slice using a Pt-Ir bipolar electrode with a stimulator (Scout, ripple neuro), while the electro-optic sensor was positioned underneath the CA1 region. The same optical setup as optical characterization was used for the optical recording and a tungsten wire microelectrode was used for the electrical recording, which was also inserted in the CA1 region.

The ground-truth electrical response was obtained from tungsten wire electrode which was connected to a neural amplifier headstage (RHD 32ch, Intan Technologies). The optical responses from the electro-optic sensor were obtained by fixing the wavelength of laser near the resonance wavelength. A template subtraction to fit the rising and falling part with exponential functions is necessary for both electrical and optical responses to show the evoked local field potentials (LFPs).

# References

1. Marblestone, A. H. *et al.* Physical principles for scalable neural recording. *Front Comput Neurosci* (2013) doi:10.3389/fncom.2013.00137.

2. Ye, Z. *et al.* Ultra-high density electrodes improve detection, yield, and cell type identification in neuronal recordings. *bioRxiv* 2023.08.23.554527 (2024) doi:10.1101/2023.08.23.554527.

3. Chamanzar, M., Denman, D. J., Blanche, T. J. & Maharbiz, M. M. Ultracompact optoflex neural probes for high-resolution electrophysiology and optogenetic stimulation. in *2015 28th IEEE International Conference on Micro Electro Mechanical Systems (MEMS)* 682–685 (IEEE, 2015). doi:10.1109/MEMSYS.2015.7051049.

4. Chamanzar, M., Denman, D. J., Blanche, T. J. & Maharbiz, M. M. Ultracompact optoflex neural probes for high-resolution electrophysiology and optogenetic stimulation. in *2015 28th IEEE International Conference on Micro Electro Mechanical Systems (MEMS)* 682–685 (IEEE, 2015). doi:10.1109/MEMSYS.2015.7051049.

5. Jun, J. J. *et al.* Fully integrated silicon probes for high-density recording of neural activity. *Nature* (2017) doi:10.1038/nature24636.

6. Steinmetz, N. A. *et al.* Neuropixels 2.0: A miniaturized high-density probe for stable, long-term brain recordings. *Science (1979)* **372**, eabf4588 (2021).

7. Angotzi, G. N. *et al.* SiNAPS: An implantable active pixel sensor CMOS-probe for simultaneous large-scale neural recordings. *Biosens Bioelectron* (2019) doi:10.1016/j.bios.2018.10.032.

8. Masvidal-Codina, E. *et al.* High-resolution mapping of infraslow cortical brain activity enabled by graphene microtransistors. *Nat Mater* (2019) doi:10.1038/s41563-018-0249-4.

9. Bonaccini Calia, A. *et al.* Full-bandwidth electrophysiology of seizures and epileptiform activity enabled by flexible graphene microtransistor depth neural probes. *Nat Nanotechnol* (2022) doi:10.1038/s41565-021-01041-9.

10. Dreier, J. P. & Reiffurth, C. The Stroke-Migraine Depolarization Continuum. *Neuron* Preprint at https://doi.org/10.1016/j.neuron.2015.04.004 (2015).

11. Lauritzen, M. *et al.* Clinical relevance of cortical spreading depression in neurological disorders: Migraine, malignant stroke, subarachnoid and intracranial hemorrhage, and traumatic brain injury. *Journal of Cerebral Blood Flow and Metabolism* Preprint at https://doi.org/10.1038/jcbfm.2010.191 (2011).

12. Bonaccini Calia, A. *et al.* Full-bandwidth electrophysiology of seizures and epileptiform activity enabled by flexible graphene microtransistor depth neural probes. *Nat Nanotechnol* (2022) doi:10.1038/s41565-021-01041-9.

13. Rodriques, S. G. *et al.* Multiplexed neural recording along a single optical fiber via optical reflectometry. *J Biomed Opt* **21**, 057003 (2016).

14. Kim, S. A. *et al.* Optical measurement of neural activity using surface plasmon resonance. **33**, 914–916 (2008).

15. Kim, S. A., Kim, S. J., Moon, H. & Jun, S. B. In vivo optical neural recording using fiber-based surface plasmon resonance. **37**, 614–616 (2012).

16. Habib, A., Zhu, X., Can, U. I. & Mclanahan, M. L. Electro-plasmonic nanoantenna : A nonfluorescent optical probe for ultrasensitive label-free detection of electrophysiological signals. (2019).

17. Zhang, J., Atay, T. & Nurmikko, A. V. Optical Detection of Brain Cell Activity Using Plasmonic Gold Nanoparticles 2009. 1–6 (2009).

18. Balch, H. B. *et al.* Graphene Electric Field Sensor Enables Single Shot Label-Free Imaging of Bioelectric Potentials. *Nano Lett* (2021) doi:10.1021/acs.nanolett.1c00543.

19. Horng, J. *et al.* Imaging electric field dynamics with graphene optoelectronics. 1–7 (2016) doi:10.1038/ncomms13704.

20. Steinmetz, N. A. *et al.* Neuropixels 2.0: A miniaturized high-density probe for stable, long-term brain recordings. *Science (1979)* **372**, (2021).

21. Lopez, C. M. *et al.* 22.7 A 966-electrode neural probe with 384 configurable channels in 0.13μm SOI CMOS. in *2016 IEEE International Solid-State Circuits Conference (ISSCC)* 392–393 (IEEE, 2016). doi:10.1109/ISSCC.2016.7418072.

22. Rastogi, S. K. *et al.* Remote nongenetic optical modulation of neuronal activity using fuzzy graphene. *Proc Natl Acad Sci U S A* (2020) doi:10.1073/pnas.1919921117.

23. Brown, M. A., Crosser, M. S., Ulibarri, A. C., Fengel, C. V & Minot, E. D. Hall Effect Measurements of the Double-Layer Capacitance of the Graphene–Electrolyte Interface. *The Journal of Physical Chemistry C* **123**, 22706–22710 (2019).

24. Lu, Y. *et al.* Ultralow Impedance Graphene Microelectrodes with High Optical Transparency for Simultaneous Deep Two-Photon Imaging in Transgenic Mice. *Adv Funct Mater* (2018) doi:10.1002/adfm.201800002.

25. Forssell, M. *et al.* Effect of focality of transcranial currents on neural responses. in *2021 10th International IEEE/EMBS Conference on Neural Engineering (NER)* 289–292 (2021). doi:10.1109/NER49283.2021.9441382.

26. Sebastião, A. M., Cunha, R. A., De Mendonça, A. & Ribeiro, J. A. Modification of adenosine modulation of synaptic transmission in the hippocampus of aged rats. *Br J Pharmacol* **131**, 1629–1634 (2000).

27. Sorianello, V. *et al.* Graphene-silicon phase modulators with gigahertz bandwidth. *Nat Photonics* (2018) doi:10.1038/s41566-017-0071-6.

28. Sorianello, V., Midrio, M. & Romagnoli, M. Design optimization of single and double layer Graphene phase modulators in SOI. *Opt Express* (2015) doi:10.1364/oe.23.006478.

29. Datta, I., Ji, X., Bhatt, G. R., Lee, B. S. & Lipson, M. Low insertion loss, graphene-based platform for loss modulation. in *Optics InfoBase Conference Papers* (2022).

30. Xia, Z. *et al.* High resolution on-chip spectroscopy based on miniaturized microdonut resonators. *Opt. Express19 (13)* 12356–12364 (2011).

31. Jayatilleka, H. *et al.* Wavelength tuning and stabilization of microring-based filters using silicon in-resonator photoconductive heaters. *Optics Express, Vol. 23, Issue 19, pp. 25084-25097* **23**, 25084–25097 (2015).

32. Campenhout, J. Van *et al.* Fabrication and characterization of CMOS-compatible integrated tungsten heaters for thermo-optic tuning in silicon photonics devices. *Optical Materials Express, Vol. 4, Issue 7, pp. 1383-1388* **4**, 1383–1388 (2014).

33. Wu, Y., Shang, H., Zheng, X. & Chu, T. Post-Processing Trimming of Silicon Photonic Devices Using Femtosecond Laser. *Nanomaterials 2023, Vol. 13, Page 1031* **13**, 1031 (2023).

34. Biryukova, V., Sharp, G. J., Klitis, C., Sorel, M. & Sorel, M. Trimming of silicon-on-insulator ring-resonators via localized laser annealing. *Optics Express, Vol. 28, Issue 8, pp. 11156-11164* **28**, 11156–11164 (2020).

35. Wu, C. *et al.* Wafer-Scale Integration of Single Layer Graphene Electro-Absorption Modulators in a 300 mm CMOS Pilot Line. *Laser Photon Rev* **17**, 2200789 (2023).

36. Sheng, Z. *et al.* A compact and low-loss MMI coupler fabricated with CMOS technology. *IEEE Photonics J* (2012) doi:10.1109/JPHOT.2012.2230320.

37. Wang, Z. *et al.* A III-V-on-Si ultra-dense comb laser. *Light Sci Appl* **6**, e16260–e16260 (2017).

38. Chang, L., Liu, S. & Bowers, J. E. Integrated optical frequency comb technologies. *Nat Photonics* **16**, 95–108 (2022).

39. Insight Photonic Solutions, Inc. Akinetic All-Semiconductor Technology. https://www.sweptlaser.com/akinetic-technology.

40. Bonesi, M. *et al.* Akinetic all-semiconductor programmable swept-source at 1550 nm and 1310 nm with centimeters coherence length. *Opt Express* (2014) doi:10.1364/oe.22.002632.

41. Park, K. S. *et al.* Phase stable swept-source optical coherence tomography with active mode-locking laser for contrast enhancements of retinal angiography. *Sci Rep* (2021) doi:10.1038/s41598-021-95982-9.

42. Twayana, K. *et al.* Frequency-comb-calibrated swept-wavelength interferometry. *Opt Express* **29**, 24363–24372 (2021).

43. Moon, S. & Chen, Z. Phase-stability optimization of swept-source optical coherence tomography. *Biomed Opt Express* (2018) doi:10.1364/boe.9.005280.

44. Song, S., Xu, J., Men, S., Shen, T. T. & Wang, R. K. Robust numerical phase stabilization for long-range swept-source optical coherence tomography. *J Biophotonics* (2017) doi:10.1002/jbio.201700034.

45. Rastogi, S. K. *et al.* Graphene Microelectrode Arrays for Electrical and Optical Measurements of Human Stem Cell-Derived Cardiomyocytes. *Cell Mol Bioeng* **11**, 407 (2018).

46. Llinás, R. & Sugimori, M. Electrophysiological properties of in vitro Purkinje cell dendrites in mammalian cerebellar slices. *J Physiol* **305**, 197–213 (1980).

47. Bischofberger, J., Engel, D., Li, L., Geiger, J. R. P. & Jonas, P. Patch-clamp recording from mossy fiber terminals in hippocampal slices. *Nat Protoc* **1**, 2075–2081 (2006).

# Supplementary Information/ Extended Data

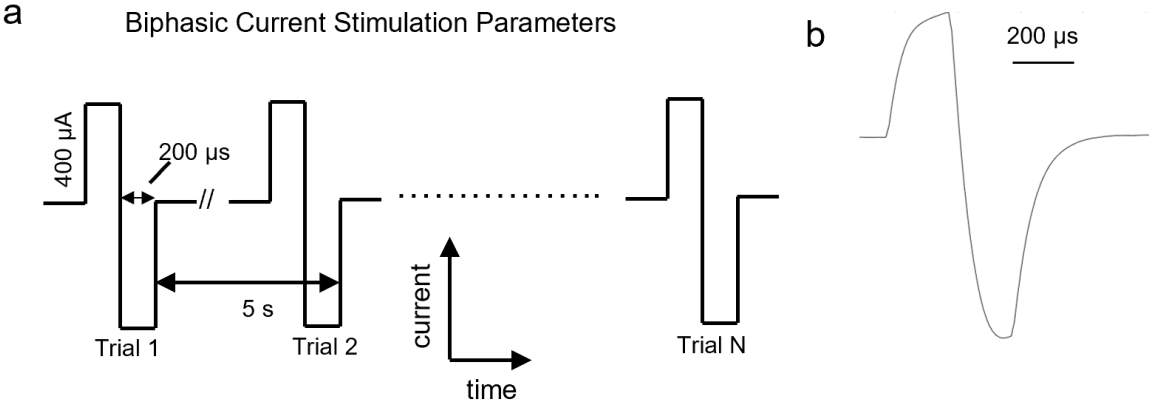

*Figure S1: (a) Stimulation parameters for biphasic current stimulation used to evoke neural response in cortical and hippocampal mouse brain slices. (b) Optically recorded biphasic stimulation artifact from the electro-optic sensor.*

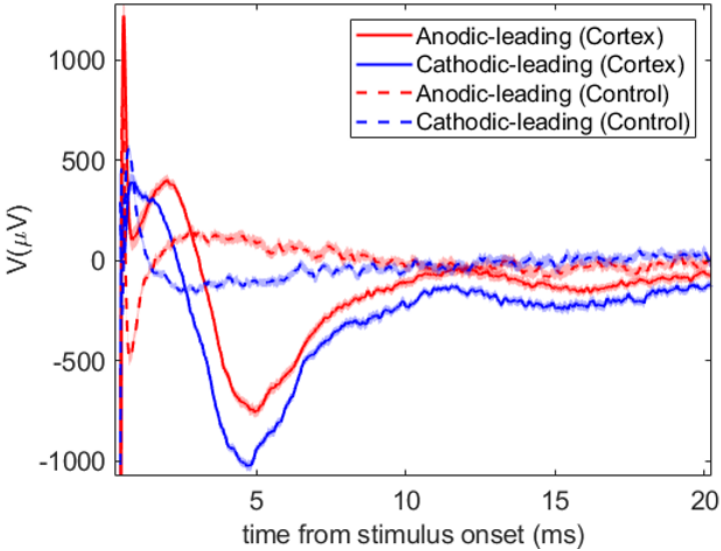

*Figure S2: Graphs illustrating the recorded responses in both live (cortex) and dead brain slices (control) subjected to stimulation with different leading polarities.*

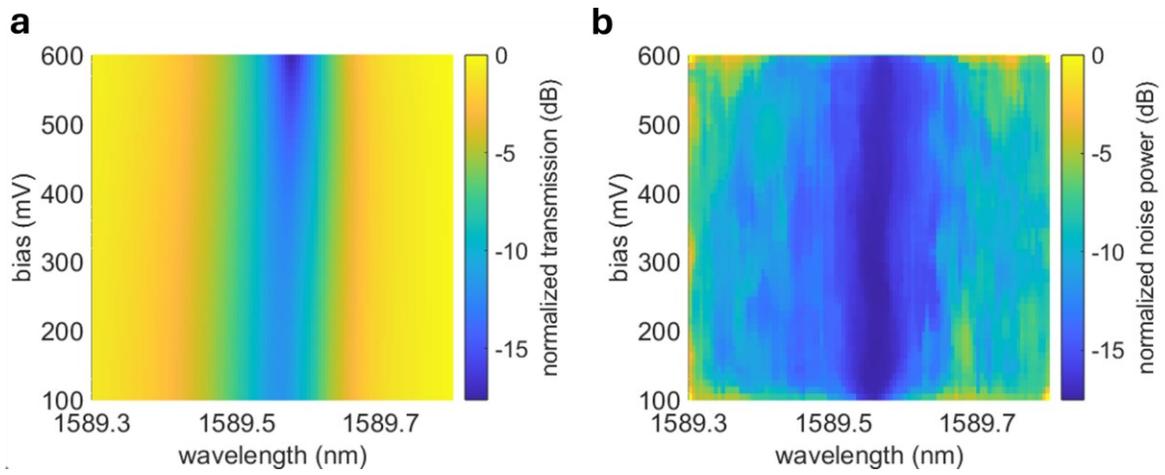

Figure S3: Measured (a)normalized transmission and (b)normalized noise power as a function of wavelength and bias

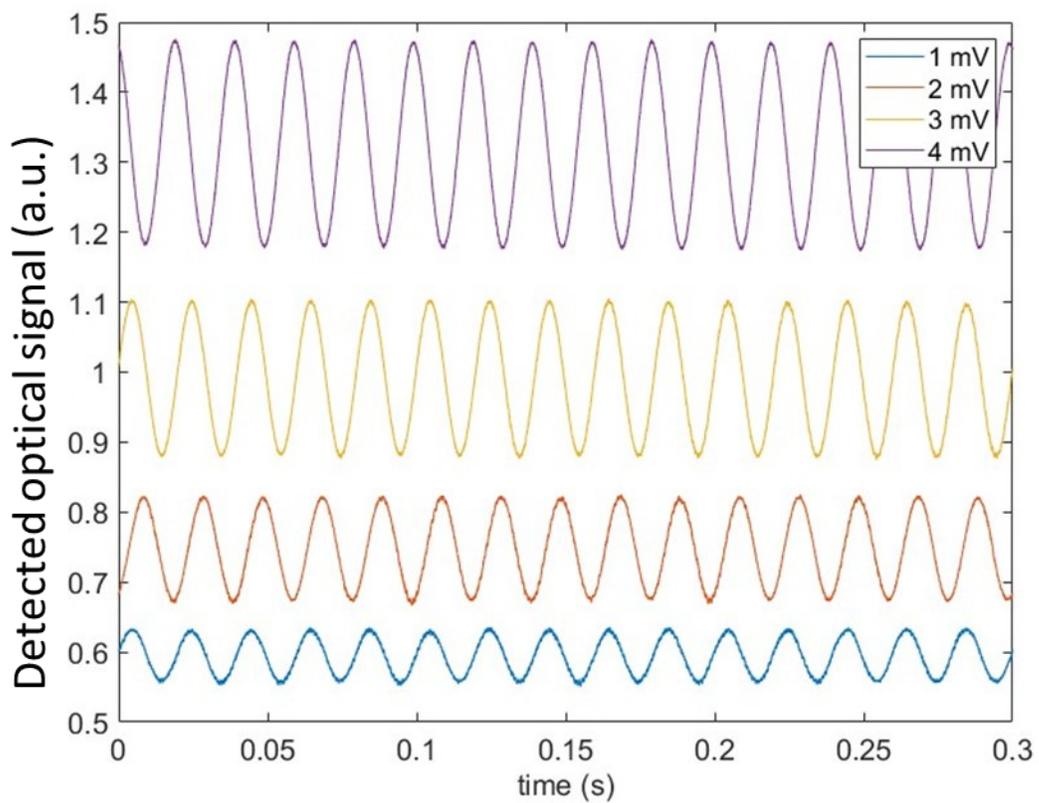

Figure S4: Optical signal detected from electro-optic sensor when modulating sinusoidal signals were applied. An SNR of 30 dB is estimated for the modulating signal with 1 mV amplitude.

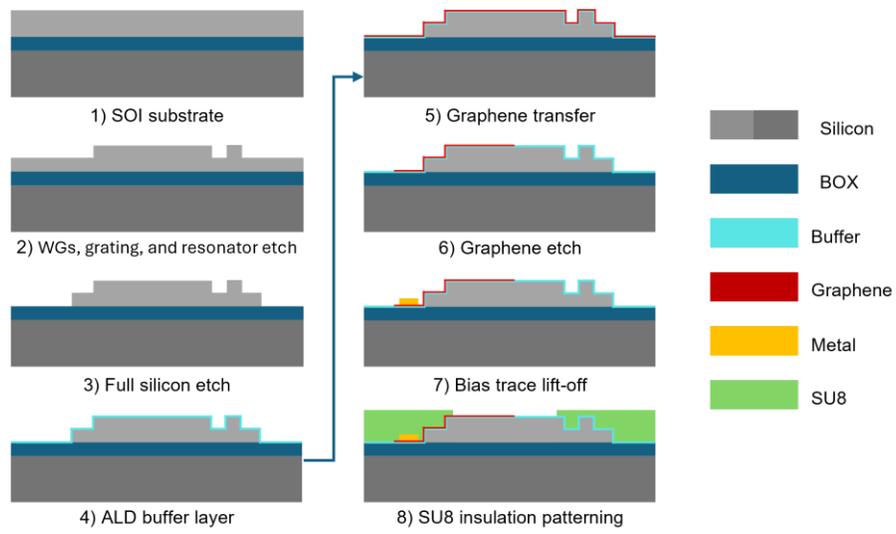

*Figure S5 Simplified fabrication process of graphene-integrated microresonator based electro-optic sensor.*